\documentclass[12pt]{article}

\usepackage[centertags]{amsmath}
\usepackage{amsfonts}
\usepackage{amssymb}
\usepackage{amsthm}
\usepackage{cite}
\usepackage{epsfig}
\usepackage{calc}

\DeclareMathOperator{\sgn}{sgn}

\begin{document}


\title{\textbf{Effective dynamics of an electrically charged string with a current}}

\author
{ P.O. Kazinski\thanks{E-mail: kpo@phys.tsu.ru}\\
\textit{Physics Faculty, Tomsk State University,}\\
\textit{Tomsk 634050, Russia} }

\maketitle

\begin{abstract}

Equations of motion for an electrically charged string with a current in an external
electromagnetic field with regard to the first correction due to the self-action are
derived. It is shown that the reparametrization invariance of the free action of the
string imposes constraints on the possible form of the current. The effective
equations of motion are obtained for an absolutely elastic charged string in the
form of a ring (circle). Equations for the external electromagnetic fields that
admit stationary states of such a ring are revealed. Solutions to the effective
equations of motion of an absolutely elastic charged ring in the absence of external
fields as well as in an external uniform magnetic field are obtained. In the latter
case, the frequency at which one can observe radiation emitted by the ring is
evaluated. A model of an absolutely nonstretchable charged string with a current is
proposed. The effective equations of motion are derived within this model, and a
class of solutions to these equations is found.

\end{abstract}

\section{Introduction}

The description of the effective dynamics of electrically charged low-dimensional
objects, such as particles, strings, and membranes, is one of traditional problems
in classical electrodynamics. The application of such models allows one to
considerably simplify the solving of the system of Maxwell--Lorentz
integrodifferential equations. For a nonrelativistic charged particle, the effective
equations of motion were obtained as early as by Lorentz \cite{Lor}. The
relativistic generalization of the Lorentz equations was derived by Dirac
\cite{Dir}. At present, the effective equations of motion are known for a point
charge in a curved background space--time \cite{DeWBr}, for a spinning particle
\cite{BU, RR}, for a massive particle in higher dimensions \cite{Kos, rrmp}, and for
a massless charged particle in the four-dimensional space--time \cite{rrmlp}. The
general scheme for the description of the self-action of electric currents in the
string models is given in \cite{Car}. In  \cite{BP}, the general theory of moving
electrically charged relativistic membranes is described.

In the present paper, we give an approximate (neglecting the effect of radiative
friction) Poincar\'{e}-invariant description of the effective dynamics of a thin
electrically charged string with a current. The importance of studying the effective
dynamics of such strings is beyond doubt because of the numerous applications, both
in practice and theoretical models, of extended charged and / or conducting objects
with negligible transverse dimensions. For instance, the effective equations of
motion obtained in Section \ref{charged string} are applied to two specific models
of strings in Sections \ref{charged loop} and \ref{non-stretched string}. In Section
\ref{charged loop} we consider the effective dynamics of an absolutely
elastic\footnote{We define an absolutely elastic string as a string that does not
significantly resist both external forces and the forces induced by its own fields.
For example, an imaginary line with a current may serve as such a string. One should
not confuse this concept with the well-known model of the Nambu--Goto string in the
limit of zero tension (see, for example, \cite{ILST}), where the string yet has its
own dynamics.} ring-shaped charged string. This model describes the dynamics of a
high-current beam of charged particles that move along a circle. In Section
\ref{non-stretched string}, we study the effective dynamics of an absolutely
nonstretchable charged string with a current. In Section \ref{charged string}, we
derive the effective equations of motion for a charged string with a current and
discuss some of their properties; in particular, in the case of a
reparametrization-invariant free action of a string, we find the generators of gauge
transformations and the constraints on the possible form of the current that flows
along the string.

We will describe a charged string within the model of an infinitely thin string. It
is well known that the self-action of such a string leads to a diverging expression
for the force of the self-action, because infinitely close points of an infinitely
thin charged string interact with infinite force. The regularization procedure,
whose physical meaning consists in ``smearing'' a singular source of the
electromagnetic field, allows one to represent the self-action force as an
asymptotic series in the regularization parameter -- the cross-section radius of the
string -- which contains one logarithmically divergent term. The smaller the
cross-section radius of the string, the greater the contribution of this divergent
term to the self-action force. For a sufficiently thin string, one may neglect other
terms of the asymptotic series to obtain effective equations of motion for a thin
charged string in the form of a system of differential equations rather than
integrodifferential equations, as in the case when all terms of the asymptotic
series are taken into account. Similar equations are obtained when describing the
effective dynamics of cosmic strings (see review \cite{HindKib}).

\section{A charged string with a current}\label{charged string}

In this section, we find the leading contribution of the self-action of an
electrically charged string with a current and derive equations of motion for the
string in an external field with regard to this correction. We show that the
requirement of the reparametrization invariance of the free action of a string
imposes constraints on the possible form of the current flowing through the string.

Suppose given a closed string $N$ with coordinates $\{\tau, \sigma\}$,
$\sigma\in[0,2\pi)$, that is embedded by a smooth mapping $x(\tau,\sigma)$ into the
Minkowski space $\mathbb{R}^{3,1}$ with coordinates $\{x^\mu\}$, $\mu=0,\ldots,3$
and the metric $\eta_{\mu\nu}=diag(1,-1, -1,-1)$. Suppose that $e(\tau,\sigma)$ is a
vector density defined on the string  $N$ that characterizes the electric current
flowing through the string. Then, from the viewpoint of an ambient space, the
current density is given by (see, for example, \cite{BP})
\begin{equation}\label{current charged string}
    j^\mu(x)=c\int{\delta^4(x-x(\tau,\sigma))e^i(\tau,\sigma)\partial_ix^\mu(\tau,\sigma) d\tau d\sigma},
\end{equation}
where $c$ is the velocity of light; it is obvious that the charge conservation law
$\partial_\mu j^\mu=0$ immediately implies $\partial_ie^i=0$. Hereupon, the Latin
indices run through the values $0,1$ and correspond to $\tau$ and $\sigma$
respectively.

Let us introduce a nondegenerate symmetric scalar product in a linear space of
$n$-forms on $\mathbb{R}^{3,1}$ as follows:
\begin{equation}
    \langle X,
    Y\rangle=n!\int\limits_{\mathbb{R}^{3,1}}{X\wedge*Y}=\int\limits_{\mathbb{R}^{3,1}}{X_{\mu_1\ldots\mu_n}Y^{\mu_1\ldots\mu_n}d^4x},
\end{equation}
here $*$ is the Hodge operator that sends $n$-forms to $(4-n)$-forms, and $\wedge$
denotes the exterior product of forms. In these terms, the action of the model in
question is expressed as
\begin{equation}\label{action charged string}
    S[A,x]=-\frac1{8\pi c}\langle A,\delta dA\rangle-\frac1{c^2}\langle j,A\rangle+S_0[x],
\end{equation}
where $d$ is the exterior differential, $\delta=*d*$, $A_\mu$ is the $4$-potential
of the electromagnetic field, and $S_0[x]$ is the action that describes the free
dynamics of the string. The equations of motion for action \eqref{action charged
string} are given by
\begin{equation}
    \delta F=-\frac{4\pi}cj,\;\;\;\;\;\frac{\delta S_0[x]}{\delta x^\mu}=\frac1{c^2}\langle\frac{\delta j[x]}{\delta
    x^\mu},A\rangle=\frac1cF_{\mu\nu}e^i\partial_ix^\nu,
\end{equation}
where $F=dA$ is the strength tensor of the electromagnetic field.

To obtain effective equations of motion of a string, we should solve the Maxwell
equations for an arbitrary configuration of the string and substitute the solutions
of these equations into the expression for the Lorentz force. This yields an
ill-defined (divergent) expression for the self-action force of the string:
\begin{equation}
   F^{rr}_\mu[x]=-\frac{4\pi}{c^3}\langle\frac{\delta j[x]}{\delta
    x^\mu},Gj[x]\rangle,
\end{equation}
where $G$ is an operator whose kernel is a retarded Green's function. Applying a
regularization procedure \cite{siss} to this expression, we obtain an asymptotic
series in the regularization parameter that contains one logarithmically divergent
term. The regularization parameter makes the sense of the cross-section radius of
the string; when this radius tends to zero, the radiation reaction force diverges.
If the cross-section radius is small but finite, then this divergent term makes the
most essential contribution to the self-action force; moreover, the smaller the
cross-section radius, the larger this contribution.

Using the formalism developed in \cite{siss} we can easily show that the
logarithmically divergent term that arises in the expression for the self-action
force can be obtained by varying the action with the Lagrangian\footnote{This result
can even be obtained without invoking the general covariant procedure, proposed in
\cite{siss} for regularizing the radiation reaction in theories with singular
sources. The leading divergent term is uniquely determined by the Poincar\'{e}
invariance and the reparametrization invariance and by the expression multiplying
the $\delta$-function in formula \eqref{current charged string}. These arguments are
frequently used for deriving leading divergent terms \cite{Wit, Car, BuoDam}.}
\begin{equation}\label{rrlagrangian}
    L^{sing}=-\frac1{c}\left.\frac{e^i\partial_ix_\mu e^j\partial_jx^\mu}{\sqrt{|h|}}\,2\ln\frac{\Lambda}{\varepsilon}\right|_{\varepsilon\rightarrow0}=-\frac1{c}\left.\frac{e^2}{\sqrt{|h|}}\,2\ln\frac{\Lambda}{\varepsilon}\right|_{\varepsilon\rightarrow0},
\end{equation}
where $e^2=e^ie^jh_{ij}$, $h_{ij}=\partial_ix_\mu\partial_jx^\mu$ is the induced
metric on the string, $h=\det h_{ij}$, the parameter $\Lambda$ characterizes the
cut-off of the integral at the upper limit (its magnitude is on the order of the
string length), and $\varepsilon$ is the cut-off parameter of the integral at the
lower limit (its magnitude is on the order of the cross-section radius of the
string).

Let us introduce a vector field $V^i=e^i/\sqrt{|h|}$ and a $1$-form
$\upsilon_i=h_{ij}V^i$. Then, neglecting the finite part of the radiation reaction
force, we obtain the following effective equations of motion of the string:
\begin{equation}\label{force covariant}
    \frac{\delta S_0[x]}{\delta
    x^\mu}=\frac\chi{c}(T^{ij}\nabla_{ij}x_\mu+\nabla_iT^{ij}\partial_jx_\mu)\sqrt{|h|}+\frac1cF_{\mu\nu}e^i\partial_ix^\nu,\;\;\;\;\;T^{ij}=(V^2h^{ij}-2V^iV^j),
\end{equation}
where $\chi=2\ln(\Lambda/\varepsilon)$ is a dimensionless constant, $F_{\mu\nu}$is
the strength tensor of the external electromagnetic field, and $\nabla_i$ is a
connection compatible with the metric $h_{ij}$. The traceless tensor $T^{ij}$
represents the density of the energy--momentum tensor corresponding to Lagrangian
\eqref{rrlagrangian}; i.e.,
\begin{equation}
  T^{sing}_{\mu\nu}=\frac{\chi}{c}\int{\delta^4(x-x(\tau,\sigma))T^{ij}\partial_ix_\mu\partial_jx_\nu\sqrt{|h|}d\tau d\sigma}.
\end{equation}
The tracelessness of the tensors $T^{ij}$ and $T^{sing}_{\mu\nu}$ follows from the
conformal invariance of Lagrangian \eqref{rrlagrangian}.

If the free action $S_0[x]$ of the string is reparametrization-invariant, then the
equations of motion \eqref{force covariant} possess a ``residual'' reparametrization
invariance, which implies that the equations are orthogonal to the vector
$V^i\partial_ix^\mu$. In addition, we have
\begin{equation}\label{energy consrevation law}
    \nabla_kT^k_i=-2i_Vd\upsilon_i=-\chi^{-1}\partial_ix^\mu F_{\mu\nu}V^k\partial_kx^\nu.
\end{equation}
In particular, in the absence of an external field, the above equality and the
charge conservation law $\nabla_iV^i=0$ imply
\begin{equation}\label{equalities on v}
    d\upsilon=0,\;\;\;\;\;\delta\upsilon=0,
\end{equation}
i.e. $\upsilon$ is a harmonic 1-form. If the closed string has no
self-intersections, Eqs. \eqref{equalities on v} are easily solved. Applying the
conformal gauge
\begin{equation}\label{conformal gauge}
  \dot{x}_\mu x'^\mu=0,\;\;\;\;\;\dot{x}_\mu\dot{x}^\mu=-x'_\mu x'^\mu,
\end{equation}
where the dot denotes the differentiation with respect to $\tau$, and the prime
denotes the differentiation with respect to $\sigma$, we obtain the following
expressions for the general solution to Eqs. \eqref{equalities on v}:
\begin{equation}\label{solution to equalities on v}
  \lambda=\lambda_0+f(\sigma+\tau)+g(\sigma-\tau),\;\;\;\;\;I=I_0-f(\sigma+\tau)+g(\sigma-\tau).
\end{equation}
Here, we used more customary notations $\lambda=e^0$ and $I=e^1$, $\lambda_0, I_0$
are arbitrary constants, and $f$ and $g$ are arbitrary $2\pi$-periodic functions. In
other words, in the absence of an external field, the energy--momentum conservation
law results in the relation between the linear density of charge and the current
\eqref{solution to equalities on v}.

Within our approximation, Eqs. \eqref{equalities on v} represent a mathematical
expression for the condition that the string is superconducting (has no resistance):
one of these equations states the charge conservation law, and the other states
that, in the absence of external fields, the time-variation of the current at a
given point of the string is attributed only to the gradient of the linear density
of charge. The fulfillment of these equations follows from the requirement that the
free action $S_0[x]$ should be reparametrization-invariant. The converse is also
true: the superconductivity conditions \eqref{equalities on v} for an arbitrary
configuration of the string imply the reparametrization invariance of its free
action.

Using the energy--momentum conservation law \eqref{energy consrevation law} we can
rewrite the equations of motion of the string in an external field as
\begin{equation}\label{effeqmotion}
    \frac{\delta S_0[x]}{\delta x^\mu}=\frac\chi{c}T^{ij}\nabla_{ij}x_\mu\sqrt{|h|}+\frac1{c}\gamma^\rho_\mu F_{\rho\nu}e^i\partial_ix^\nu,
\end{equation}
where $\gamma_{\mu\nu}=\eta_{\mu\nu}-h^{ij}\partial_ix_\mu\partial_jx_\nu$. Thus, if
the free action of an electrically charged string with a current is
reparametrization-invariant, then its effective dynamics in an external
electromagnetic field are described by the system of equations \eqref{energy
consrevation law}, \eqref{effeqmotion}.

When the contribution of the singular term is sufficiently large, i.e., when the
string is sufficiently thin ($\chi\gg1$) and either the current flowing through it
or the linear density of charge are large, one can neglect the left-hand sides of
Eqs. \eqref{force covariant}; in this case, the free effective dynamics of the
string are completely determined by the leading contribution to the self-action
force of the charged string. We say that such a string is absolutely elastic because
its internal structure does not appreciably resist an action.

When the current density increases further, one can also neglect the effect of the
external field; then, the effective dynamics of the string are described by the
equation
\begin{equation}\label{free effeqmotion}
  T^{ij}\nabla_{ij}x_\mu=0,
\end{equation}
provided that $\upsilon$ is a harmonic $1$-form.

Further, we will solve the system of equations \eqref{energy consrevation law},
\eqref{effeqmotion} for the model of a ring-shaped absolutely elastic string in an
external electromagnetic field and consider the model of an absolutely
nonstretchable charged string with a current; for the latter model, we will derive
the effective equations of motion and obtain certain particular solutions.

\section{A charged ring}\label{charged loop}

As we pointed out in the Introduction, the model of an absolutely elastic charged
string describes a high-current beam of charged particles; therefore, it is
worthwhile to consider its effective dynamics in an external electromagnetic field.
In this section, we consider the effective dynamics of an absolutely elastic charged
string in the form of a ring (a circle). Further, we derive equations for external
electromagnetic fields that admit stationary states of such a ring. Then, we find
solutions to free equations of motion and solve the equations of motion of a
uniformly charged ring in an external uniform magnetic field. The last mentioned
model describes the behavior of a high-current beam of charged particles in a
synchrotron.

Consider a gauge that is convenient for further calculations. Introduce coordinates
$\{\tau,\sigma\}$ so that the vector density $e^i$ is straightened in these
coordinates; i.e., it has the form $e=(1,0)$. Let us show that such coordinates can
be introduced without changing the coordinate $\tau$.

Suppose that, in the original coordinates $\{t,l\}$ the vector density $e$ has
components $(\lambda, I)$, then, in the coordinates $\{\tau,\sigma\}$ we obtain
\begin{equation}
  e^0=\frac{\dot\tau\lambda+\tau'I}{\dot\tau\sigma'-\tau'\dot\sigma},\;\;\;\;\;e^1=\frac{\dot\sigma\lambda+\sigma'I}{\dot\tau\sigma'-\tau'\dot\sigma},
\end{equation}
here, the dots and primes denote the differentiation with respect to $t$ and $l$
respectively. Setting $e^0=1$, $e^1=0$ and $\dot\tau=1$, $\tau'=0$, $\sigma'\neq0$
we obtain the following relations for $\sigma$
\begin{equation}\label{equations on sigma}
  \dot\sigma=-I,\;\;\;\;\;\sigma'=\lambda.
\end{equation}
provided that $\lambda\neq0$ at this point. Equations \eqref{equations on sigma} are
integrable by virtue of the charge conservation law. For example, if
$dl=\sqrt{\partial_\sigma\mathbf{x}\partial_\sigma\mathbf{x}}\,d\sigma$ is a length
element of the string, then the linear density of charge is represented as
\begin{equation}
  |\lambda|=(\partial_\sigma\mathbf{x}\partial_\sigma\mathbf{x})^{-1/2}.
\end{equation}
In other words, the coordinate $\sigma$ counts the charge on the string. Next, we
will assume that $\lambda\neq0$ throughout the string..

In addition to the above gauge, we require that
\begin{equation}
  x^0(\tau,\sigma)=c\tau\;\;\Rightarrow\;\;\tau=t.
\end{equation}
Then, the metric induced on the world sheet of the ring
\begin{equation}
  x^0=ct,\;\;\;\;\;\mathbf{x}(t,\sigma)=r(t)(\cos{\varphi(t,\sigma)},\sin{\varphi(t,\sigma)},0),
\end{equation}
and its inverse are given by
\begin{equation}
  h_{ij}=\left[%
\begin{array}{cc}
  c^2-\dot{r}^2-r^2\dot{\varphi}^2 & -r^2\dot{\varphi}\varphi' \\
  -r^2\dot{\varphi}\varphi' & -r^2\varphi'^2 \\
\end{array}%
\right],\;\;\;\;\;
  h^{ij}=\left[%
\begin{array}{cc}
  \dfrac1{c^2-\dot{r}^2} & -\dfrac{\dot\varphi}{\varphi'(c^2-\dot{r}^2)} \\
  -\dfrac{\dot\varphi}{\varphi'(c^2-\dot{r}^2)} & -\dfrac{c^2-\dot{r}^2-r^2\dot{\varphi}^2}{r^2\varphi'^2(c^2-\dot{r}^2)} \\
\end{array}%
\right].
\end{equation}
Hereupon, the prime denotes the differentiation with respect to $\sigma$. The
determinant of the induced metric is equal to $h=-r^2\varphi'^2(c^2-\dot{r}^2)$. The
functions $\dot{\varphi}(t,\sigma)$ and $\varphi'(t,\sigma)$ are smooth and
$Q$-periodic in the variable $\sigma$, where $Q=const$ s the total charge of the
ring. The linear density of charge is equal to $\lambda=(r\varphi')^{-1}$; here, we
matched the signs of $\lambda$ and $\varphi'$. The fundamental harmonic of
$\varphi'(t,\sigma)$ with respect to $\sigma$ is equal to $2\pi Q^{-1}$; in
particular, $\varphi'=2\pi Q^{-1}$ if the ring is uniformly charged.

The absolute elasticity of a string implies that the free action $S_0[x]$ of the
string is identically zero. In this case, Eqs. \eqref{effeqmotion} are rewritten as
\begin{equation}\label{effeqmotion for loop}
  T^{ij}\nabla_{ij}x_\mu\sqrt{|h|}=-\gamma^\rho_\mu F_{\rho\nu}e^i\partial_ix^\nu,
\end{equation}
where the external field is redefined as $F_{\mu\nu}\rightarrow\chi F_{\mu\nu}$.
Throughout this section, the expressions for the electromagnetic fields will contain
$\chi^{-1}$. We will also assume that the external field is cylindrically symmetric
and that $E_z=H_r=H_\varphi=0$ where, as usual, the subscripts indicate the
projections of a vector onto an appropriate unit vector. Then, Eqs. \eqref{energy
consrevation law} and \eqref{effeqmotion for loop} are equivalent to the following
two equations:
\begin{equation}\label{eqmotion for loop general}
\begin{split}
  \ddot{\varphi}-2\frac{\dot{\varphi}'\dot{\varphi}}{\varphi'}-\varphi''\frac{c^2-\dot{r}^2-r^2\dot{\varphi}^2}{r^2\varphi'^2}+\frac{\dot{r}\dot{\varphi}}{r}\frac{c^2-\dot{r}^2+r\ddot{r}}{c^2-\dot{r}^2}=\frac{\varphi'}2(c^2-\dot{r}^2)^{1/2}(cE_\varphi-H_z\dot{r}),\\
  \dfrac{(c^2-\dot{r}^2+r^2\dot{\varphi}^2)(c^2-\dot{r}^2-r\ddot{r})}{r^2\varphi'(c^2-\dot{r}^2)^{3/2}}=\frac{r\dot{\varphi}}c(E_\varphi\dot{r}-cH_z)-\frac{E_r}c(c^2-\dot{r}^2).
\end{split}
\end{equation}
In particular, the first equation implies the equation that defines the variation
law for the effective angular momentum of an absolutely elastic charged ring:
\begin{equation}\label{conservation law for loop}
    \frac{d}{dt}\left[\int\limits_0^Q{d\sigma\frac{r\dot{\varphi}}{\sqrt{c^2-\dot{r}^2}}}\right]=\pi
    r(cE_\varphi-H_z\dot{r}).
\end{equation}

Let us consider the stationary states of a charged ring in the external field; i.e.,
let us set $r(t)=const$ in Eqs. \eqref{eqmotion for loop general}. Then, we obtain
\begin{equation}\label{phi at stationary points}
  \begin{split}
    \ddot{\varphi}-2\frac{\dot{\varphi}'\dot{\varphi}}{\varphi'}-\varphi''\frac{c^2-r^2\dot{\varphi}^2}{r^2\varphi'^2}=\frac{c^2\varphi'}2E_\varphi,\;\;\;\;\;
  \dfrac{c^2+r^2\dot{\varphi}^2}{r^2c\varphi'}=-r\dot{\varphi}H_z-cE_r,
  \end{split}
\end{equation}
where the external fields are, generally speaking, certain functions of $t$. Formula
\eqref{conservation law for loop} can be rewritten as
\begin{equation}
    \frac{d}{dt}\int\limits_0^Q{d\sigma\dot{\varphi}}=\pi c^2E_\varphi.
\end{equation}
The second equation in \eqref{phi at stationary points} implies that
\begin{equation}\label{velocity angle}
  \frac{r\dot\varphi}{c}=-\frac{r^2\varphi'H_z}2\pm\sqrt{\frac{r^4\varphi'^2H_z^2}{4}-(1+r^2\varphi'E_r)}.
\end{equation}

\begin{figure}[t]
\centering\epsfig{file=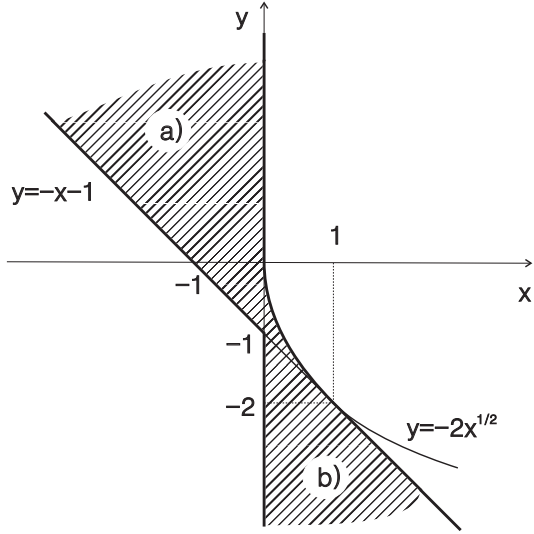, width=7cm} \caption{{\footnotesize The region
where stationary states of a positively ($\varphi'>0$) charged ring may exist is
crosshatched. $y=r^2\varphi'H_z$, $x=1+r^2\varphi'E_r$ and $\dot\varphi\geq0$. The
region $a)$, which is bounded from below by the straight line $y=-x-1$, corresponds
to the situation when Eq. \eqref{velocity angle} is taken with sign $+$, and the
region $b)$, which is bounded by the axis $y$ and the curve $y=-2x^{1/2}$,
corresponds to the branch with sign $-$. In the overlap of regions $a)$ and $b)$
there may exist stationary states of the ring that have different angular velocities
for the same values of the external field, charge, and ring radius.}}\label{plot for
stable config}
\end{figure}

The requirements that the radicand be nonnegative and that the velocity of the
string be less than the velocity of light impose constraints on the fields $H_z$ and
$E_r$, under which stable states of the ring may exist. These requirements are
illustrated graphically in Fig. \ref{plot for stable config}. For example, if there
is no electric field and a ring of radius $r$ is uniformly charged, then the
magnetic field can hold this ring only if
\begin{equation}\label{stationary point for H const}
  |H_z(r)|\geq\frac{|Q|}{\pi r^2},
\end{equation}
where $Q$ is the total charge of the ring.

Equations \eqref{phi at stationary points} are rather complicated in the general
case; therefore, we restrict the analysis to a uniformly charged ring
($\varphi''=0$) for $E_r=0$. Then, we have
\begin{equation}\label{phi stationary}
  \ddot\varphi=\frac{c^2\varphi'}2E_\varphi,\;\;\;\;\;\varphi'=\frac{2\pi}{Q},\;\;\;\;\;
  \frac{r\dot\varphi}{c}=-\frac{\pi r^2H_z}{Q}-\sqrt{\frac{\pi^2r^4H_z^2}{Q^2}-1},
\end{equation}
where we used the fact that the equality $\varphi''=0$  implies the equality
$\dot\varphi'=0$. The substitution of the expression for $\dot\varphi$ into the
equation for $\ddot\varphi$ in \eqref{phi stationary} yields equations for the
fields that admit such stationary configurations.

For example, in the nonrelativistic limit $r^2\dot\varphi^2\ll c^2$, we obtain the
following solution to \eqref{phi stationary}:
\begin{equation}\label{phi stationary nonrel}
  \frac{r\dot\varphi(t)}c=-\frac{Q}{2\pi
  r^2H_z(t)},\;\;\;\;\;\varphi'(\sigma)=\frac{2\pi}{Q},\;\;\;\;\;2\pi^2r^3c^2E_\varphi(t)=Q^2\frac{\dot{H}_z(t)}{H_z^2(t)},
\end{equation}
whereas, in the ultrarelativistic limit $r^2\dot\varphi^2\approx c^2$, Eqs.
\eqref{phi stationary} lead to the equalities
\begin{equation}\label{phi stationary ultrarel}
  \frac{r\dot\varphi(t)}c=-\frac{Q}{\pi r^2H_z(t)},\;\;\;\;\;\varphi'(\sigma)=\frac{2\pi}{Q},\;\;\;\;\;\dot{H}_z(t)=-c\frac{E_\varphi(t)}{r}(1+\frac{\pi r^2}{Q}H_z(t)).
\end{equation}
Thus, a uniformly charged ring does not change its radius only if the external
fields satisfy Eqs. \eqref{phi stationary} (or Eqs. \eqref{phi stationary nonrel}
and \eqref{phi stationary ultrarel} in the nonrelativistic and ultrarelativistic
cases, respectively), provided, of course, that $E_r=0$.

Now, we proceed to solving the dynamical equations \eqref{eqmotion for loop
general}. Consider the case when there are no external fields. The solution of the
second equation in \eqref{eqmotion for loop general} yields
\begin{equation}\label{solution for r free}
  r(t)=\sqrt{\rho^2+c^2(t+\tau)^2},\;\;\;\;\;\rho=r(0)\sqrt{1-\frac{\dot{r}^2(0)}{c^2}},\;\;\;\;\;\tau=\frac{r(0)\dot{r}(0)}{c^2}.
\end{equation}
Then, the first equation takes the form
\begin{equation}\label{equation on phi free}
  \ddot\varphi-2\frac{\dot\varphi'\dot\varphi}{\varphi'}-\frac{\varphi''}{\varphi'^2}(\frac{c^2\rho^2}{r^4}-\dot\varphi^2)+2\frac{\dot{r}\dot\varphi}{r}=0,
\end{equation}
whence
\begin{equation}
    \frac{d}{dt}\int\limits_0^Q{d\sigma r^2\dot{\varphi}}=0.
\end{equation}
If, in addition, we require that $\dot\varphi'=0$, which physically means that all
points of the ring rotate with the same velocity, then we obtain the solution
\begin{equation}
  r^2(t)\dot\varphi(t)=r^2(0)\dot\varphi(0),\;\;\;\;\;\varphi'(\sigma)=\frac{2\pi}{Q}.
\end{equation}
As expected, the equation for $\varphi$ represents the angular momentum conservation
law.

Solution \eqref{solution for r free} shows that, after a certain period of time, the
ring will expand with a velocity close to the velocity of light; therefore, it is
worthwhile to consider the ultrarelativistic limit of Eq. \eqref{equation on phi
free}; i.e., it is worthwhile to require that
$c^2-\dot{r}^2-r^2\dot\varphi^2\approx0$. In this case, we have the following
conservation law:
\begin{equation}
  \frac{r^2(t)\dot\varphi(t,\sigma)}{\varphi'^2(t,\sigma)}=\frac{r^2(0)\dot\varphi(0,\sigma)}{\varphi'^2(0,\sigma)}.
\end{equation}
This equation can be solved by the method of characteristics (see, for example,
\cite{Arn}). In a particular case when
\[
    \dot\varphi(0,\sigma)/\varphi'^2(0,\sigma)=\lambda^2(0,\sigma)r^2(0)\dot\varphi(0,\sigma)=b=const,
\]
i.e., when the linear density of the effective angular momentum is the same at all
points of the string, we obtain
\begin{equation}
\begin{split}
  \dot\varphi(t,\sigma)&=\frac{br^2(0)}{r^2(t)}\varphi'^2(0,\sigma_0(t,\sigma)),\\
  \sigma&=\sigma_0(t,\sigma)-\frac{2br^2(0)}{\rho
  c}\left(\arctan{\frac{c(t+\tau)}{\rho}}-\arctan{\frac{c\tau}{\rho}}\right)\varphi'(0,\sigma_0(t,\sigma)).
\end{split}
\end{equation}
The above equation should be considered as an equation for $\sigma_0(t,\sigma)$ for
a certain prescribed $Q$-periodic function $\varphi'(0,\sigma_0)$ whose fundamental
harmonic is equal to $2\pi Q^{-1}$.

To conclude this section, consider the effective dynamics of a charged ring all of
whose points move with the same angular velocity ($\dot{\varphi}'=0$) in a uniform
magnetic field $H_z=const$. In this case, from \eqref{eqmotion for loop general} and
\eqref{conservation law for loop} we obtain the system of equations
\begin{equation}\label{equation at const H}
  \frac{(c^2-\dot{r}^2+r^2\dot{\varphi}^2)(c^2-\dot{r}^2-r\ddot{r})}{r^2\varphi'(c^2-\dot{r}^2)^{3/2}}=-r\dot{\varphi}H_z,\;\;\;\;\;\frac{r\dot{\varphi}}{\sqrt{c^2-\dot{r}^2}}+\frac{\pi
  H_z}{2Q}r^2=M,
\end{equation}
where $M$ is a certain constant defined by the initial data. The first equation
implies, in particular, that $\varphi'=2\pi Q^{-1}$. Substituting the expression for
$r\dot{\varphi}$ from the second equation into the first, we obtain an equation for
the function $r(t)$ alone, which has the form
\begin{equation}
    \sqrt{c^2-\upsilon^2}=k(1+(M-(\pi H_z/2Q)r^2)^2)/r,
\end{equation}
where $\upsilon(r(t))\equiv\dot{r}(t)$ and $k$ is an integration constant. Let us
express the equations of motion in dimensionless variables. Introduce $r_0=|\pi
H_z/2Q|^{-1/2}$ and redefine $r$ and $t$ as $r\rightarrow r_0r$ and $t\rightarrow
t\,r_0/c$.  For example, the velocity in these coordinates is measured in the units
of the velocity of light. Then, the equations of motion of a charged ring are
expressed as
\begin{equation}\label{eqmotion for loop preintegrated}
    \frac{\upsilon^2}2+\frac12\left[k^2\frac{(1+L^2)^2}{r^2}-1\right]=0,\;\;\;\;\;\dot{\varphi}=k\frac{L(1+L^2)}{r^2},
\end{equation}
where $L=M-\sgn{(H_z/Q)}r^2$.

\begin{figure}[t]
\centering\epsfig{file=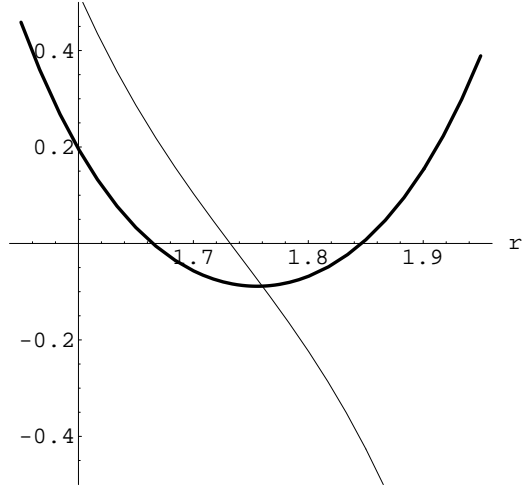, width=7cm} \caption{{\footnotesize The functions of
the potential $U(r)$ (heavy line) and the angular velocity $r\dot{\varphi}$ (thin
line). The diagrams correspond to the initial data $r(0)\approx1.844$,
$\upsilon(0)=0.1$, $M=3$ and $\sgn{(H_z/Q)}=1$. The potential attains its minimum at
the point $r_{ext}\approx1.755$.}}\label{plot for U}
\end{figure}

The first equation in \eqref{eqmotion for loop preintegrated} resembles the equation
of motion of a particle of unit mass with zero total energy in the potential field
$U(r)=\left[k^2r^{-2}(1+L^2)^2-1\right]/2$ the only difference between these
equations is that the form of $U(r)$ depends on the initial data $r(0)$,
$\upsilon(0)$ and $\dot\varphi(0)$ (see Fig. \ref{plot for U}). The potential has a
single extremum at the point
\begin{equation}
    r_{ext}^2=\frac13(\sgn{(\frac{H_z}{Q})}M+\sqrt{4M^2+3}),
\end{equation}
and indefinitely increases as $r\rightarrow0$ and $r\rightarrow+\infty$ ; therefore,
for any initial data, the system will oscillate about the equilibrium point
$r_{ext}$. Note that the minimal value of $r_{ext}$ is equal to $\sqrt2/2$, which
agrees with the results of the previous analysis of the stationary states
\eqref{stationary point for H const} of a charged ring.

We can evaluate the ratio of the oscillation frequency $\omega$ of the charged ring
in the neighborhood of the equilibrium point $r_{ext}$ to the mean angular frequency
$\omega_0$ of its rotation. In the harmonic approximation, the first frequency is
defined by $U''(r_{ext})$, and the second, by $|\dot{\varphi}(r_{ext})|$; hence, we
have
\begin{equation}\label{fraction formula}
  \frac{\omega}{\omega_0}\approx\frac1r\frac{(4r^2-\sqrt{4r^4-1})^{1/2}}{(2r^2-\sqrt{4r^4-1})^{3/2}},
\end{equation}
where $r=r_{ext}$. This ratio uniformly increases from $2$ to infinity as $r_{ext}$
increases; for large values of $r_{ext}$ it increases as $8\sqrt2 r_{ext}^3$. For
example, a threefold increase in the linear dimensions of a synchrotron leads to a
ninefold increase in the oscillation frequency for the same mean angular velocities
of the high-current beam of particles, other characteristics, such as $H_z$, $Q$ and
$\chi$, remaining constant.

Thus, an absolutely elastic uniformly charged ring in a uniform magnetic field
oscillates about the equilibrium point  $r_{ext}$, $\dot\varphi_{ext}$, according to
Eqs. \eqref{eqmotion for loop preintegrated} with frequency $\omega$; the ratio of
this frequency to the angular frequency of the ring is defined by \eqref{fraction
formula}. Therefore, we can expect that, when the energy inflow compensates the
energy losses, a high-current beam of charged particles in a synchrotron will also
produce radiation at this frequency, in addition to the well-known synchrotron
radiation. For example, if one could separate these two types of radiation by
certain characteristics and measure the ratio $\omega/\omega_0$, then one would
determine the equilibrium position $r_{ext}$ in the units of $r_0$ by formula
\eqref{fraction formula}.

Among the disadvantages of the model of an absolutely elastic charged string as
applied to the description of a high-current beam of charged particles is the fact
that this model does not take into account the radiation reaction due to the
synchrotron radiation, which becomes significant at large angular velocities.

\section{An absolutely nonstretchable string}\label{non-stretched string}

In this section, we consider the dynamics of a thin, absolutely nonstretchable
charged string\footnote{Possibly, a more customary term for the model of an
absolutely nonstretchable string is a perfect weightless thread, which comes from
mechanics. A detailed description of the theory of an absolutely flexible thread can
be found, for example, in \cite{Merk}.} with a current with regard to the
first-order correction due to the self-action. We derive equations of motion for
such a string, investigate its stationary states in the absence of an external
electromagnetic field, and find a class of solutions to the equations of motion for
a uncharged string with a current and for a uniformly charged string.

Suppose given a closed string $N$ with coordinates $\{\tau, \sigma\}$,
$\sigma\in[0,L)$, that is embedded into the Minkowski space $\mathbb{R}^{3,1}$ by a
smooth mapping $x(\tau)$. In relativistic mechanics, the concept of
nonstretchability makes sense only in a certain distinguished frame of reference.
Let us introduce a $4$-vector $n^\mu$, $n^2=1$, that characterizes such a frame of
reference; then, the action that describes the free dynamics of an absolutely
nonstretchable string is given by
\begin{equation}\label{action non-stretched string}
  S_0[x,\varkappa,\kappa^0,\kappa^1]=\int\limits_N{d\tau d\sigma\left(\varkappa(h_{00}+h)+\kappa^0(n^\mu\dot{x}_\mu-c)+\kappa^1n^\mu x'_\mu\right)},
\end{equation}
where $\varkappa$, $\kappa^0$ and $\kappa^1$ are Lagrange multipliers of the
constraints that guarantee the nonstretchability of the string. It is obvious that
action \eqref{action non-stretched string} is not invariant under the change of
coordinates $\{\tau,\sigma\}$, hence, there are no constraints \eqref{energy
consrevation law} on the form of the linear density of charge and the current in
this case.

The effective equations of motion \eqref{force covariant} or a thin absolutely
nonstretchable charged string with a current in the distinguished frame of reference
$n^\mu=(1,0,0,0)$ are expressed as
\begin{equation}\label{eqmotion for non-stretched string}
  \begin{split}
    -&\partial_i(L^{ij}\partial_jx_\mu+\kappa^i\delta^0_\mu)=\frac1cF_{\mu\nu}e^i\partial_ix^\nu,\;\;\;\;\;L^{ij}\equiv\frac{\chi}c\frac{h_{kl}e^ke^lh^{ij}-2e^ie^j}{\sqrt{|h|}}+2\varkappa(\delta^i_0\delta^j_0+hh^{ij}),\\
    h&_{00}=-h,\;\;\;\;\;\dot x^0=c,\;\;\;\;\;x'^0=0.
  \end{split}
\end{equation}
The last three equations in \eqref{eqmotion for non-stretched string} represent the
condition of ``relativistic nonstretchability''. Indeed, the equation $h_{00}=-h$
has the form
\begin{equation}\label{non-stretch requirement}
  \mathbf{x}'^2=1-\frac{(\dot{\mathbf{x}}\mathbf{x}')^2 }{c^2-\dot{\mathbf{x}}^2 },
\end{equation}
which immediately implies that the coordinate $\sigma$ is a natural parameter on the
string with a correction for relativistic contraction. Since $\sigma$ ranges from
$0$ to $L$, the fulfillment of equality \eqref{non-stretch requirement} at any point
of the string automatically implies that the string of length $L$ is nonstretchable.

Since we consider a closed string, all functions entering Eq. \eqref{eqmotion for
non-stretched string} must be periodic in $\sigma$. For an open string with free
ends, the periodicity condition is replaced by the equality
\begin{equation}
    L^{1i}\partial_ix_\mu+\kappa^1\delta^0_\mu=0,
\end{equation}
at the ends of the string.

We will consider the effective dynamics of a closed string in the absence of
external electromagnetic fields. The unknown fields $\kappa^i$ can be obtained from
the first two equations in \eqref{eqmotion for non-stretched string} by setting
$\mu=0$. As a result, we are left with four equations for four unknown functions
$\mathbf{x}(t,\sigma)$ and $\varkappa(t,\sigma)$:
\begin{equation}\label{eqmotion for non-stretched string refined}
    L^{ij}\partial_{ij}\mathbf{x}+\partial_iL^{ij}\partial_j\mathbf{x}=0,\;\;\;\;\;\mathbf{x}'^2=1-\frac{(\dot{\mathbf{x}}\mathbf{x}')^2 }{c^2-\dot{\mathbf{x}}^2 }.
\end{equation}
The consistency condition for this system yields an equation for
$\varkappa(t,\sigma)$. The physical meaning of the field $\varkappa(t,\sigma)$ is
that it compensates for the forces that stretch (contract) the string.

Let us find the stationary configurations of the string that are consistent with
Eqs. \eqref{eqmotion for non-stretched string refined}, i.e. set
$\dot{\mathbf{x}}\equiv0$ in the equations of motion. We can easily show that, in
this case, the equations of motion are reduced to the system\footnote{Henceforth, we
redefine the Lagrangian multiplier $\varkappa$ as follows
$\varkappa\rightarrow-\varkappa\chi/c$.}
\begin{equation}
  2\varkappa'+\frac{(\lambda^2)'}{c}+2\frac{\lambda\dot{I}}{c^3}=0,\;\;\;\;\;2\varkappa+\frac{\lambda^2}{c}+\frac{I^2}{c^3}=0,\;\;\;\;\;\mathbf{x}'^2=1.
\end{equation}
Taking into account the charge conservation law, $\dot{\lambda}+I'=0$, we have
\begin{equation}
  \varkappa=-\frac{c^2\lambda^2+I^2}{2c^3},\;\;\;\;\;\dot{(\lambda
  I)}=0,\;\;\;\;\;\mathbf{x}'^2=1.
\end{equation}
Thus, if the product of the charge density multiplied by the current is independent
of time and all points of the string are at rest at the initial moment, then there
exists $\varkappa(t,\sigma)$ such that the string retains its initial configuration.
The question concerning the stability of such solutions remains open.

Next, consider the nonrelativistic dynamics of a string described by Eqs.
\eqref{eqmotion for non-stretched string refined} in the absence of the current,
$I=0$ (charged dielectric), as well as for $\lambda=0$ (uncharged conductor). The
nonrelativistic limit is understood in the following sense. Let $l$ be a
characteristic scale of variation of the field $\mathbf{x}(t,\sigma)$, for example,
the length of the string; then, we formally define the order of smallness as
follows:
\begin{equation}
    [\dot{\mathbf{x}}/c]\ll1,\;\;\;\;\;[l^k\partial^{k+1}_\sigma\mathbf{x}]=[l\dot{\mathbf{x}}'/c]=[l\ddot{\mathbf{x}}/c^2]=1.
\end{equation}
In this case, the order of smallness of $\varkappa$ is determined from Eqs.
\eqref{eqmotion for non-stretched string refined}.

For a charged dielectric, when $I=0$ and $\dot{\lambda}=0$, we obtain the following
system of equations from \eqref{eqmotion for non-stretched string refined} in the
nonrelativistic limit:
\begin{equation}\label{eqmotion for non-stretched diel}
   \frac{\lambda^2}{c^2}\ddot{\mathbf{x}}+\Lambda\mathbf{x}''+\left[\Lambda'+\frac{\ddot{\mathbf{x}}\mathbf{x}'}{c^2}\Lambda\right]\mathbf{x}'=0,\;\;\;\;\;\mathbf{x}'^2=1,
\end{equation}
where $\Lambda=2c\varkappa+\lambda^2$. Note that these equations are invariant under
the Galilean transformations. Equations \eqref{eqmotion for non-stretched diel} can
be resolved for the higher derivative only if $\Lambda\neq-\lambda^2$. In this case,
we have
\begin{equation}\label{eqmotion for non-stretched diel 1}
    \frac{\lambda^2}{c^2}\ddot{\mathbf{x}}+\Lambda\mathbf{x}''+\frac{\lambda^2\Lambda'}{\lambda^2+\Lambda}\mathbf{x}'=0,\;\;\;\;\;\mathbf{x}'^2=1.
\end{equation}
The consistency condition for system \eqref{eqmotion for non-stretched diel 1} leads
to the following equation for $\Lambda(t,\sigma)$
\begin{equation}\label{equation on lambda}
    \left(\frac{\Lambda'}{\lambda^2+\Lambda}\right)'-\frac{\Lambda\mathbf{x}''^2}{\lambda^2}-\frac{\dot{\mathbf{x}}'^2}{c^2}=0.
\end{equation}
Thus, fixing $\Lambda(t,0)$ and $\Lambda'(t,0)$, as well as $\mathbf{x}(0,\sigma)$
and $\dot{\mathbf{x}}(0,\sigma)$, subject to the conditions
$\mathbf{x}'(0,\sigma)\mathbf{x}'(0,\sigma)=1$ and
$\dot{\mathbf{x}}'(0,\sigma)\mathbf{x}'(0,\sigma)=0$, we can construct a unique
solution $\mathbf{x}(t,\sigma)$ that satisfies Eqs. \eqref{eqmotion for
non-stretched diel 1}. Recall that all the functions must be $L$-periodic in the
variable $\sigma$. This condition imposes constraints on the boundary conditions for
the function $\Lambda(t,\sigma)$.

\begin{figure}[t]
\centering\epsfig{figure=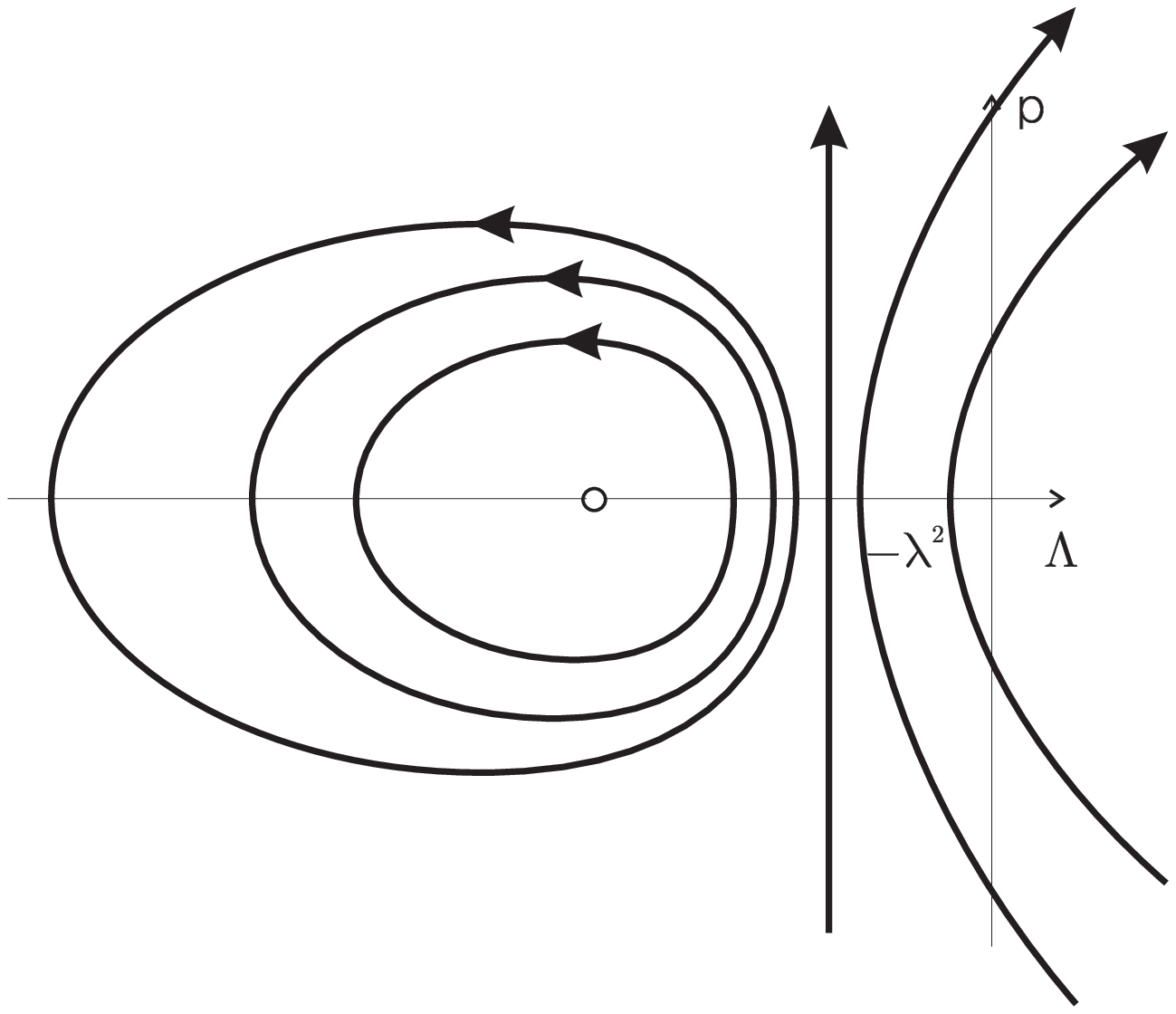, width=0.32\linewidth} \hfill
\epsfig{figure=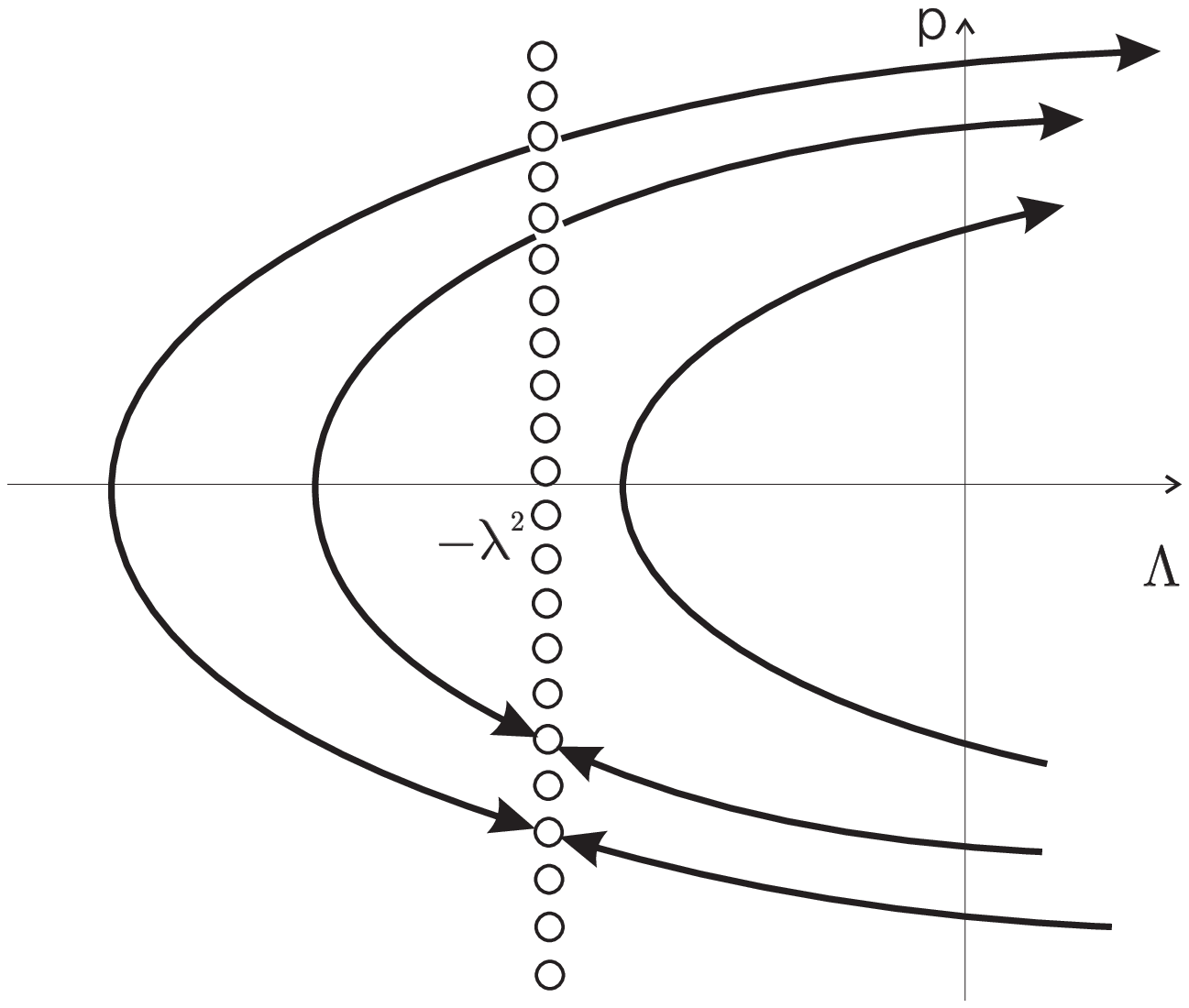, width=0.32\linewidth} \hfill
\epsfig{figure=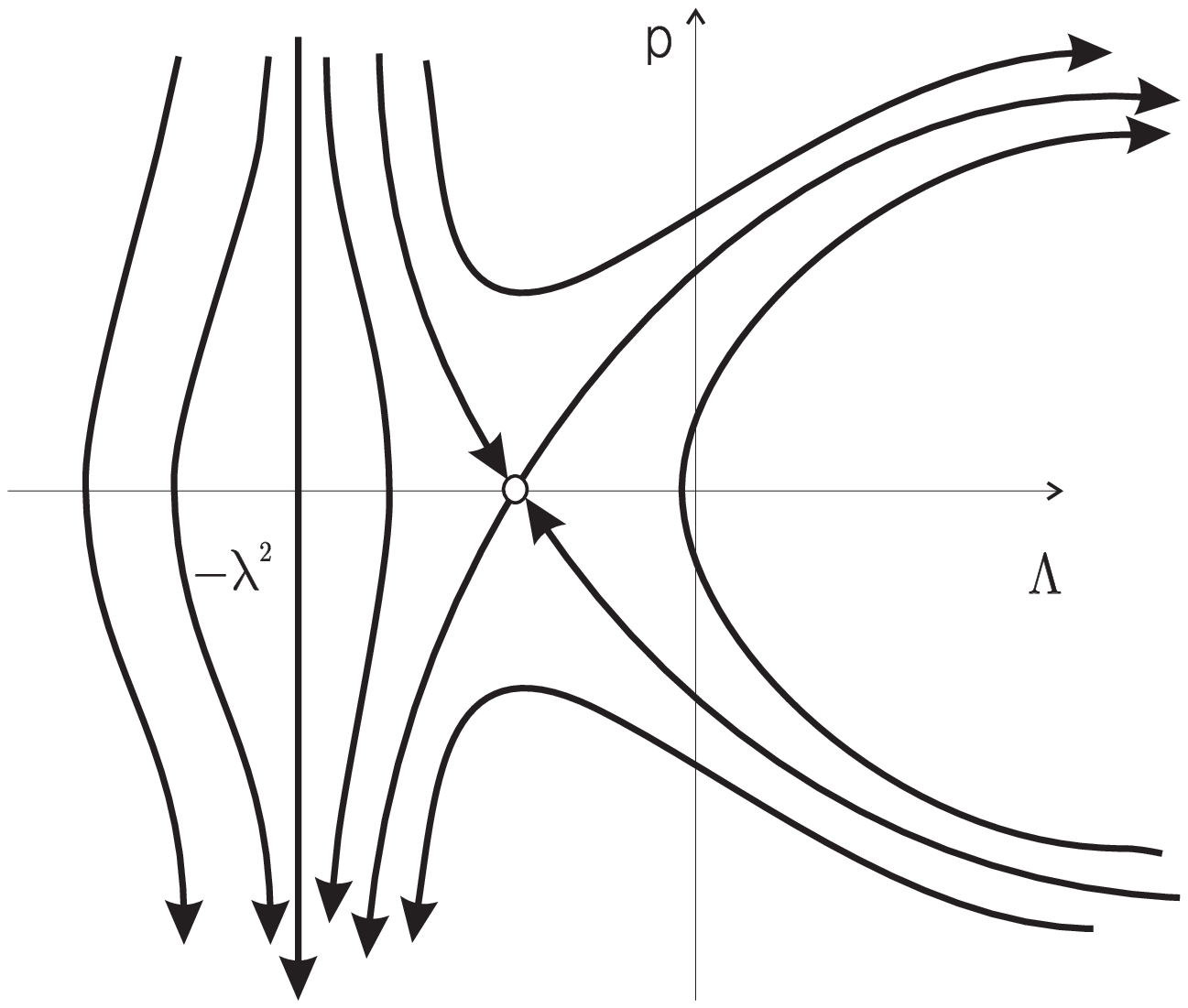, width=0.32\linewidth} \caption{{\footnotesize The
instantaneous phase portrait for Eq. \eqref{equation on lambda}, here
$p=\Lambda'/(\lambda^2+\Lambda)$. The circles indicate the singular points of the
vector field. At the left $\dot{\mathbf{x}}'^2/c^2>\mathbf{x}''^2$(this case can be
referred to as the case of small curvature of the string), at the middle
$\dot{\mathbf{x}}'^2/c^2=\mathbf{x}''^2$, and at the right
$\dot{\mathbf{x}}'^2/c^2<\mathbf{x}''^2$.}}\label{phase-plane portret}
\end{figure}

One can draw an instantaneous phase portrait for Eq. \eqref{equation on lambda} (see
Fig. \ref{phase-plane portret}). In particular, this portrait shows that, if
$\Lambda(t,0)\neq-\lambda^2$, then $\Lambda(t,\sigma)\neq-\lambda^2$,
$\forall\sigma\in[0,L)$, i.e. the quantity $(1+\Lambda/\lambda^2)$ does not change
its sign.

For a uniformly charged string, $\lambda=const$, Eqs. \eqref{eqmotion for
non-stretched diel 1} and \eqref{equation on lambda} are rewritten as
\begin{equation}\label{eqmotion for non-stretched diel lambda con}
    \frac{\ddot{\mathbf{x}}}{c^2}-(1-\sgn{(1+\frac{\Lambda}{\lambda^2})}e^\Theta)\mathbf{x}''+\Theta'\mathbf{x}'=0,\;\;\;\;\;\Theta''+(1-\sgn{(1+\frac{\Lambda}{\lambda^2})}e^\Theta)\mathbf{x}''^2-\frac{\dot{\mathbf{x}}'^2}{c^2}=0,
\end{equation}
where $\Theta=\ln{|1+\Lambda/\lambda^2|}$.

Using similar arguments for an uncharged conductor with a current, $\lambda=0$ and
$I'=0$, we obtain the following equations for $\varkappa\neq0$ (this condition is an
analogue of the inequality $\Lambda\neq-\lambda^2$, which arises in the case of a
charged dielectric)
\begin{equation}\label{eqmotion for non-stretched cur}
    \frac{\ddot{\mathbf{x}}}{c^2}+(1+\sgn{(\varkappa)}e^\Theta)\mathbf{x}''+\Theta'\mathbf{x}'=0,\;\;\;\;\;\Theta''-(1+\sgn{(\varkappa)}e^\Theta)\mathbf{x}''^2-\frac{\dot{\mathbf{x}}'^2}{c^2}=0,
\end{equation}
where $\Theta=\ln{|2c^3\varkappa/I^2|}$. The instantaneous phase portrait for the
second equation in \eqref{eqmotion for non-stretched cur} has always the same form
(with obvious redefinitions) as the instantaneous phase portrait in the case of a
weakly curved charged dielectric (Fig. \ref{phase-plane portret} at the left).
Therefore, periodic (in $\sigma$) solutions to the second equation in
\eqref{eqmotion for non-stretched cur} may only exist when $\varkappa(t,0)<0$.

Let us find certain particular solutions to the equations obtained. Setting
$\Theta'(t,\sigma)=0$ in Eqs. \eqref{eqmotion for non-stretched diel lambda con} or
\eqref{eqmotion for non-stretched cur}, we obtain the system
\begin{equation}\label{eqmotion for non-stretched at crit point}
    \ddot{\mathbf{x}}-u^2(t)\mathbf{x}''=0,\;\;\;\;\;\dot{\mathbf{x}}'^2=u^2(t)\mathbf{x}''^2,\;\;\;\;\;\mathbf{x}'^2=1.
\end{equation}
In this case, the dynamics of a uniformly charged dielectric and an uncharged
conductor are described by the same system of equations. Let us simplify the
situation by assuming that $u(t)=const$. Then, there exists a class of solutions of
the form
\begin{equation}\label{solution to non-str string}
    \mathbf{x}(t,\sigma)=\mathbf{V}t+\mathbf{x}_0(\sigma+ut),
\end{equation}
where $\mathbf{V}$ is a certain constant vector and $\mathbf{x}_0(\sigma)$,
$\mathbf{x}_0'^2=1$ defines the initial configuration of the string. Substituting
the general solution of the wave equation into the remaining two equations, one can
show that \eqref{solution to non-str string} provides the only possible solutions to
Eqs. \eqref{eqmotion for non-stretched at crit point} for $u=const$. Solutions
describe a string that ``flows'' along itself with the velocity $u$.

\section{Concluding remarks}

We have investigated the effective dynamics of a thin electrically charged string
with a current with regard to the leading contribution of the self-action. This
approximation has allowed us to describe the effective dynamics of a string in an
external electromagnetic field in the form of second-order partial differential
equations and obtain their exact solutions for certain simple models of a string in
the electromagnetic field of a special form.

We have not analyzed the question concerning the stability of the solutions
obtained. This problem may become one of possible directions of further
investigations. Another direction of research may be the study of radiation
characteristics of an absolutely elastic charged ring (a high-current beam of
charged particles) in an external uniform magnetic field; the existence of such a
radiation was discussed at the end of Section \ref{charged loop}. Moreover, it would
be interesting to find other solutions to the effective equations of motion of a
string or to carry out a numerical analysis of these equations.

\section*{Acknowledgments}

I am grateful to I.V. Gorbunov and A.A. Sharapov for carefully reading the draft of
this manuscript and for their constructive criticism. This work was supported by the
RFBR grant no. 03-02-17657, the grant for Support of Russian Scientific Schools
NSh-1743.2003.2 and the grant of Federal Educational Agency no. A04-2.9-740. Author
appreciates financial support from the Dynasty Foundation and International Center
for Fundamental Physics in Moscow.


\end{document}